\documentclass[fleqn,10pt]{wlscirep}
\usepackage[utf8]{inputenc}
\usepackage[T1]{fontenc}
\usepackage{tabu}
\usepackage{comment}

\title{Border Effect of Complex Network: An analysis on the cooperation network of movie stars across different regions }

\author[1,2,3]{Wen Zhou}
\author[3,4]{Dong Zhang}
\author[5]{Chen Chen}
\author[2,6,*]{Dianbo Liu}

\affil[1]{School of Ethnology and Sociology,Yunnan University
No.2, Cuihu North Road,Wuhua District,Kunming 650091,P.R.China}

\affil[2]{Harvard University, Cambridge, MA,USA 02138
}

\affil[3]{School of Sociology and Anthropology,Sun-Yat-Sen University. 
No. 135, Xingang Xi Road,Haizhu District, Guangzhou 510275,P.R.China}

\affil[4]{School of Social and Public Administration, ChongqingTechnology and Business University.
No. 19, xuefu Road, ,Nan'an District,Chongqing 400067,P. R. China}

\affil[5]{International College,Jiangxi University of Finance and Economics, Nanchang,China,330000}

\affil[6]{Computer Science and Artificial Intelligence Laboratory,Massachusetts Institute of Technology, Cambridge, USA, MA 02139}

\affil[*]{corresponding.author: dianbo@mit.edu and  zwen03@163.com }

\begin{abstract}
People always carry out social activities within certain geographic space. The geographic boundary likes national boundaries and provincial boundaries functions as dividing policies, distributing administrative resources, defining the scope of application of laws and regulations or even the boundary of cultural region, producing fundamental influence on people’s lifestyle. Network analysis is a powerful tool to abstract complex system. In this article, we used network method to understand changes of cooperation network in movie industry and effect of geographical boundary. 
\end{abstract}
\begin{document}

\flushbottom
\maketitle
%
%
\thispagestyle{empty}

\section*{ Introduction}
People always carry out social activities within certain geographic space. The geographic boundary likes national boundaries and provincial boundaries functions as dividing policies, distributing administrative resources, defining the scope of application of laws and regulations or even the boundary of cultural region, producing fundamental influence on people’s lifestyle (Newman,2006\cite{1}). Previous studies adopt summarized data and analyze the influence of boundary from the perspectives of inter-regional trade, travel, technology diffusion and economic coordinated development, etc and therefore form a powerful refutation to the idea of  the disappearance of geographical boundaries as well as distance (Mccallum, 1995 \cite{2}; Li \& Xu, 2006\cite{3}; 
Helliwell, 1997\cite{5}; Hazledine, 2009\cite{6}; Yu et al, 2012\cite{7}). 
In recent years, with the introduction of large-scale mobile phone location data, communication data and social network data, researchers have provided more fine-grained evidence for boundary effect from the view of complex network analysis and proved the influence of practical administrative boundary on human interaction from several aspects.(Expert et al, 2011\cite{8}; Grauwin et al, 2017\cite{9}; Hawelka et al, 2014\cite{10}; Ratti et al, 2010\cite{11}; Rinzivillo et al, 2012\cite{12}; Szell et al, 2012\cite{13}; Thiemann et al, 2010\cite{14};).
\par
Although previous employment of complex network data and research methods brings exciting change to this field, there are still several defects as follows: 1) the influence degree of geographic boundary varies with different interaction (Kang et al, 2013\cite{15}; Liu et al, 2014\cite{16}; Shi et al,  2016\cite{17}; Wang et al, 2018\cite{18}; Goldenberg \& Levy, 2009\cite{19}; Gao et al, 2013\cite{20}). Most available studies on complex network are from the perspectives of migration network and communication network and seldom involve other network. So we can’t know whether other interaction is influenced by geographic boundary likewise. 2) Under the tendency of constant expansion of technology and deep-going globalization, the judgment on the disappearance of geographic boundary is in line with the trend (Cairncross, 1997\cite{21}; Ohmae, 2015\cite{22}; Matthew et al, 2004\cite{23}; Walther \& Bernard, 2013\cite{24};Liu \& Albergante, 2017\cite{31}). However, previous studies on network mostly employ sectional data which only demonstrate the profound influence of geographic boundary on interaction but cannot judge its development tendency. 3) Under the topic of regional integration, it is urgent to break the limit of administrative boundary and realize coordinated development. Previous studies focus on the model construction and prove the influence of administrative boundary on interaction through different models. But there is no in-depth analysis on how to construct the inter-regional network, so these studies cannot provide guidance for realizing regional integration. 
\\
\\
This paper adopts the network data on the cooperation of stars in Mainland -Hongkong-Taiwan (“Mainland” is short for “Mainland China”) from 1900 to 2010, and analyzes the influence of administrative boundary on interaction from the perspective of movie star cooperation for the purpose of proving new evidence for the study in this field. Hongkong’s Return occurred in the study period and a series of policies and measures were adopted to promote the exchange and coordinated development of two places. Thus, this paper takes it as the natural experiment and takes the cooperation network of movie stars in Mainland and Taiwan in the same period as the control group to analyze the influence of the change of regional policies and political environment on complex network. Finally, this paper employs the model analysis of the number of movie stars’ cross-regional cooperation and discusses how the policies influence the development of cross-regional cooperation network. This study offers the study on geographic boundary and interpersonal interaction new and sound evidence. The in-depth analysis on policy influence can help understand how the cross-regional interpersonal interaction is influenced by policy and social environment, providing theoretical guidance for the development of regional integration.

\section*{ Materials and Methods}

\subsection*{ Data}
The data in this paper comes from the "Chinese Star Cooperative Database" constructed by the authors with on-line data fetched via Python. The cooperative relationship between stars is constructed by star’s co-starring in movies and TV plays. The data is acquired through three steps: 1) the initial Chinese-star base is sourced from ``Movie{\_}1905 (http://www.1905.com/)'', which is a website directly under directly subordinate to the State Administration of Press, Publication, Radio, Film and Television of China, and the largest provider of Chinese films. A total of 2692 film and TV stars have been sourced from the Stars’ Cooperation Database; 2) Since part of the research period of this paper was before year 1997, some of the older generation of movie stars who no longer participate in performance at present and some less famous stars are involved, so the database does not have all coverage. To expand the coverage, the authors have pulled information from the Baidu Encyclopedia of 2692 Chinese stars, including all the films and TV dramas they ever starred as well as the casts of these films and TV dramas. By then, a total of 10357 stars were covered in the database; and 3) In order to further expand the coverage and prevent omissions, the researchers followed the above method to carry out a second round of expansion, grasped the Baidu Encyclopedia of 10357 movie stars (some of the stars did not have Baidu Encyclopedia webpage at this point), obtained all their starred movies and teleplays, and got the list of a new round of movie stars. Finally, a total of 26729 stars have been covered in the database, including a lot of foreign stars from South Korea, Japan, the United States and so on. The number of Chinese and foreign films and TV drams covered is 45482. Based on these, 484804 pairs of cooperation among stars came into shape.
 \\ 

 \noindent In order to study the influence of the HongKong's return on the cooperative network of movie stars, this paper divides the research period into five periods: 1990-1993,1994-1997, 1998-2001, 2002-2005, and 2006-2009. The first two periods are used to analyze the situation and trend before HongKong's return, and the latter three periods are used to analyze the situation and trend after the return. To better reflect the real effect of the policy, this paper takes the cooperative relationships between stars from Mainland China and Taiwan as the reference group. At last, a total of 4585 stars are included in the data, and 3297 of them come from Mainland, 617 of them come from Taiwan, and 617 of them come from HongKong. There are 117,715 cooperation ties. The data of different periods is shown as follows:
\begin{table}[!ht]
\centering
\begin{tabu}to \textwidth{X[-1cm]X[-1cm]X[1cm]X[1cm]X[-1cm]X[-1cm]X[-1cm]}
\toprule
Year& 
Number of stars& 
Proportion of Hongkong stars& 
Proportion of Taiwan stars& 
Number of Edge& 
Average centrality& 
Clustering coefficient \\
\midrule
1990--1993& 
1413& 
26.82& 
12.31& 
7745& 
10.96& 
0.37 \\
1994--1997& 
1761& 
24.41& 
12.09& 
9894& 
11.24& 
0.36 \\
1998--2001& 
2276& 
19.33& 
13.00& 
15112& 
13.28& 
0.32 \\
2002--2005& 
3079& 
14.45& 
13.90& 
26214& 
17.03& 
0.32 \\
2006--2009& 
3727& 
12.24& 
13.69& 
35882& 
19.26& 
0.29 \\
\bottomrule
\end{tabu}
\caption{~~}
\label{tab1}
\end{table}
\\
\\
\noindent It can be seen from the data in the table that the number of movie stars acting the leading role tripled from Year 1990 to 2009, and the number of Taiwan stars acting the leading role also rose by nearly three times from 175 during 1990-1993 to 510 during 2006-2009, so the proportion remained unchanged. The number of Mainland stars acting the leading role soared by over four times, from 860 during 1990-1993 to 2761 during 2006-2009, thus the proportion increased from 60\% in 1990 to 74\% during 2006-2009. The number of Hongkong stars grew slowly from 379 during 1990-1993 to 456 during 2006-2009, so the proportion of Hongkong stars among all the stars dropped from 26.82\% during 1990-1993 to 12.24\% during 2006-2009.
\\ 

\noindent Fig.~\ref{fig1} below shows a cooperative network of Mainland-Hongkong-Taiwan movie stars during 1990-1993. In order to display the results more clearly, the figure takes degree of greater than 27 as the filter condition. Therefore, all the nodes denote super stars in that years. There is a total of 131 nodes and 1785 edges in the network, the circle size denotes the quantity of cooperation, and the connection line represents the cooperation between two stars. The quantity of cooperation of Yam Tat-wah(3059), Stephen Chow(5420), Carol Cheng(5332), Maggie Cheung(4940),  Leung Chiu Wai(2107), Anthony Perry(1377) and Liu Chiang(2352) are the largest, who are basically Hongkong stars.

\begin{figure}[!ht]
\centering
\includegraphics[width=.4\linewidth]{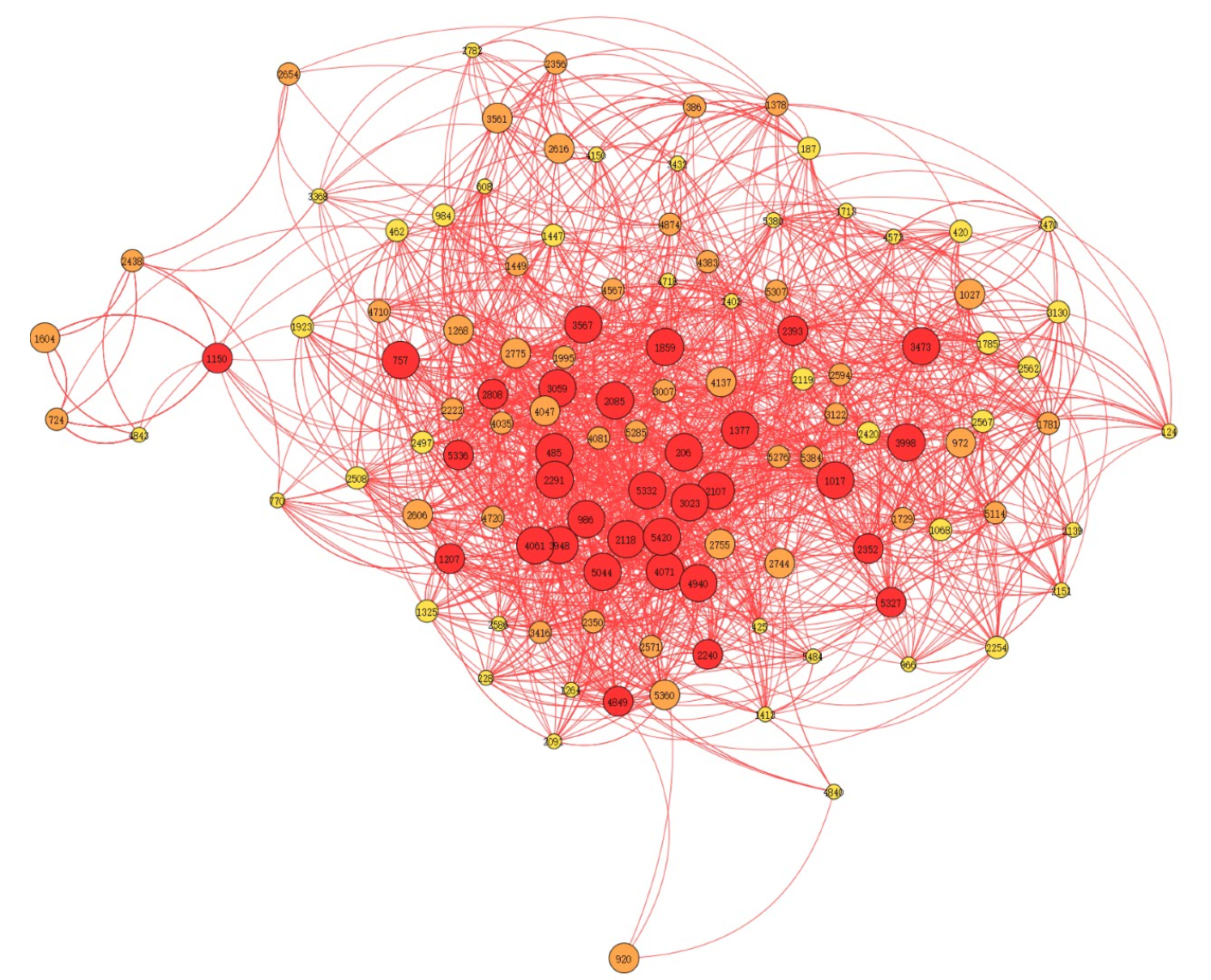}
\caption{A Cooperative Network of Mainland-Taiwan-Hongkong Movie Stars during 1990-1993 \\Fitter: degree>27 notes:131 edgers:1785}
\label{fig1}
\end{figure}

\noindent Fig.~\ref{fig2} below presents a cooperative network of Mainland-Hongkong-Taiwan movie stars during 2006-2009. The network has screened out 210 movie stars with degree of greater than 63 as the filter condition, and there are a total of 2,170 edges. It can be seen from the figure that, not merely Anthony Perry(1377), Louis Koo(948), Nicholas Tse(4267), Fong Chung Sun(757), Eric Tsang(206) and other Hongkong stars have the biggest degree in the network. Baoguo Chen(232), Guoli Zhang(4845), Guoqiang Tang(3441), Bingbing Fan(721) and other Mainland stars also become the core of the network. It can be seen by comparing Fig.1 with Fig.2 that, the cooperative network of Mainland and Hongkong stars has undergone tremendous changes during the five research periods. This paper will display these changes by constructing analysis variables and analysis methods.

\begin{figure}[!ht]
\centering
\includegraphics[width=.4\linewidth]{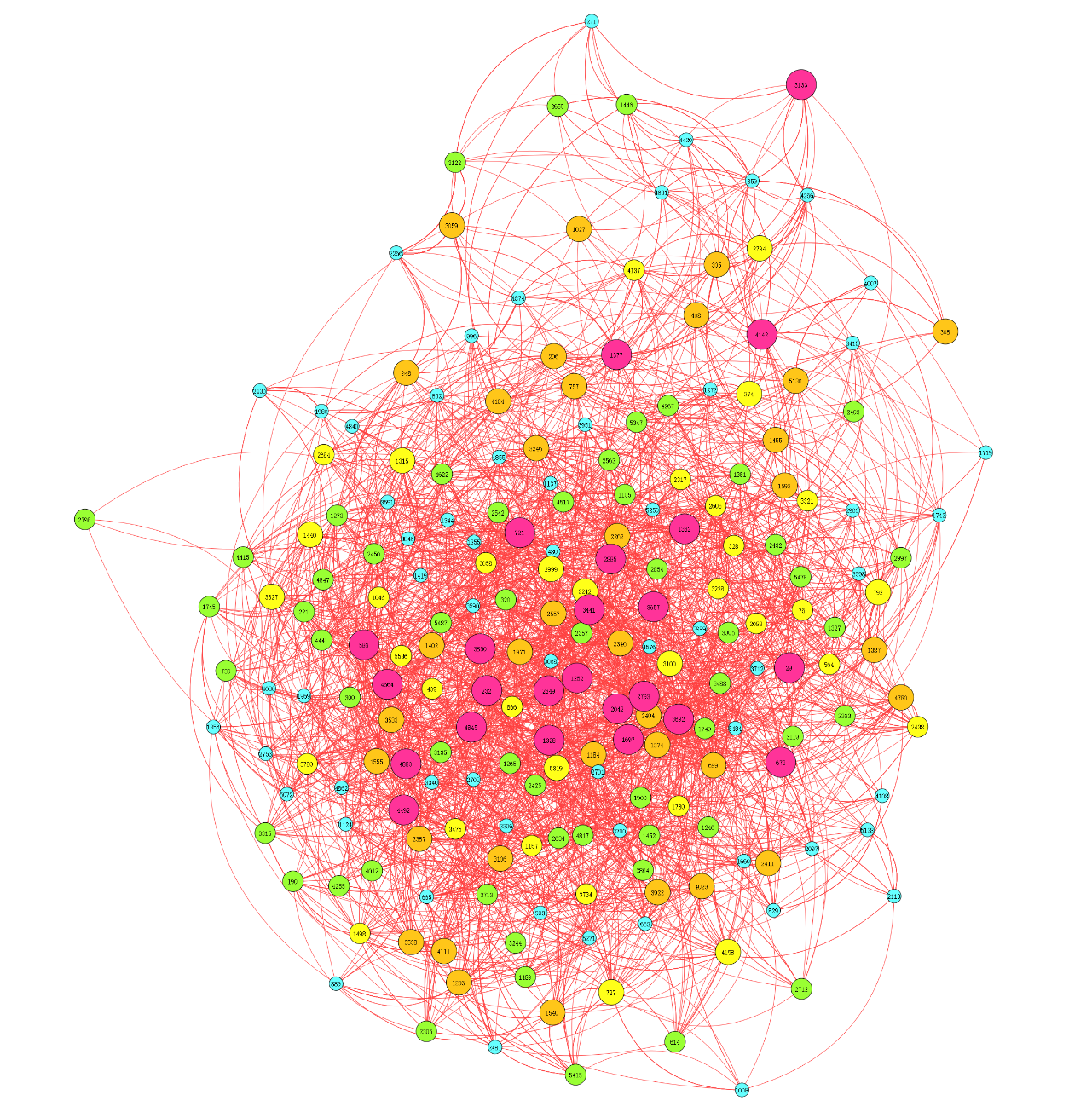}
\caption{A Cooperative Network of Mainland-Taiwan-Hongkong Movie Stars during 2006-2009 \\Fitter: degree>63 notes:210 edgers:2170}
\label{fig2}
\end{figure}

\subsection*{ Analysis Variables}
\noindent This paper studies whether the cooperation between movie stars is affected by the heterogeneity of stars’ geographical belonging, as well as the influence of the change of policy environment on cross-regional cooperation. In other words, whether the heterogeneity of stars’ geographical belonging exerts a significant impact on star cooperation and whether such impact changed around the time of Hongkong’s return are studied. The dependent variable in this paper is whether a cooperative relationship is formed between two nodes using the adjacency matrixes of all stars during different periods. The main independent variables include the stars' geographical belonging, which is indicated by birthplace and native place obtained from the Baidu Encyclopedia. The second independent variable is the popularity degree of the stars, which is measured by the stars' total cooperation quantity. Since the stars’ popularity degree can change, this paper takes their cooperation quantity in the previous period as a measurement index of their popularity degree in the present period; the third independent variable is the stars’ qualification measured by the time when they first starred a movie or TV play, the earlier the time is, the higher their seniority will be. The control variables in this paper include the age of the stars obtained from the Baidu Encyclopedia, of which some missing information is searched and complemented by artificial network, and whether the stars have cross-regional cooperation experience in the previous period is judged and generated by the database.

\subsection*{ Analysis Methods}
Table 1 indicates that the density of the stars’ cross-regional cooperative network is affected by the development trend of the star scale in the two regions. How to effectively measure the probability of cross-regional cooperation and calculate the changes in year? This paper uses two methods to measure the cross-regional stars’ cooperative network. The first method refers to Newman's community structure in networks to construct the cross-regional star cooperation index, which can calculate the probability of cross-regional stars’ cooperation by comparing with the randomised network without controlling other factors\cite{25}. The second method adopts Exponential random graph models (``ERGMs'', for short), which analyzes the impact of homogeneity of the stars' geographical belonging on the cooperation probability after controlling indexes such as the network structural factors and the node attributes including the stars' age, time of entering the profession, popularity degree and so on. Next, the two methods are introduced in detail.

\subsubsection*{ Cross-regional Stars Cooperation Index:The First Method}
\noindent\textbf{Randomised Network Model:} It is another common network analysis method to finitely and randomly disorganize the existing target network. This method randomly selects a pair of links $(a,b)$ and $(c,d)$ from the target network and changes them to links $(a,c)$ and $(b,d)$. $a,b,c,d$ are all nodes in the network (Stars). Using this method, the target network can be randomly disorganized while the integral structure of the network and the number of connections of each node can be maintained. In this paper, 2N times of random disorganizations on the stars’ cooperative network of each year are conducted, and N represents the number of all the links in the star cooperative network (Liu \& Albergante 2017 \cite{32};Liu et al. \cite{33}). The finitely disorganized network is used as the expected cooperation network of each year, and the calculation formula for the cross-regional cooperation index of each year is as follows: $\frac{\text{Observation of cooperation}(O)}{\text{Expected quantity of cooperation}(E)}$.
\\
\\
\subsubsection*{ Exponential random graph models:The Second Method}
\textbf{Exponential random graph models:} ERGMs is a common analysis model in network analysis. Similar to Generalized Linear Models (GLMs), its dependent variable is the probability of the formation of connection between two nodes. However, as the non-independence hypothesis in network analysis is not valid, ERGMs is a Logit model using Markov Chain Monte Carlo (MCMC) method (Frank \& Strauss, 1986\cite{26}; Goodreau SM, 2009\cite{27}; Wasserman \& Pattison, 1996\cite{28}; Handcock, 2003\cite{29}). By means of MCMC method, ERGMs can select the model that maximizes the generation of the network actually observed (Snijders, 2002\cite{30}). After nearly 40 years of development since 1980s, ERGMs has become an effective tool for investigating the theory of network social sciences. The probability of the observed network connections in ERGMs can be denoted as:
\begin{equation}
\label{eq1}
P(\mathbf{Y}=y|\mathbf{X})=\exp [\theta^T g(y,\mathbf{X})]/k(\theta)
\end{equation}

\noindent Where Y represents all the possible connections in the network, y is the actual or observed connections in the network, $X$ denotes the matrix of node attributes, and $g(y,\mathbf{X})$ represents the network statistics, which includes edge attribute, node attribute and network structure.These network statistics constitute the independent variables of the ERGMs. According to these independent variables, the model can predict the probability of connection formation between nodes. $\theta$ is the coefficient of the network statistics,and K is a normalization constant introduced to make the sum of the probabilities of the formation of connections between all the network statistics equal to 1.
\\
\\
\noindent ERGMs can be directly in analogy with GLMs. The dependent variable in the above equation is the probability of formation of connection between two nodes. $P_{ij}$ represents the probability of the existence of connection between two nodes, $1-P_{ij}$ denotes no connection between two nodes, and the log odds (Odds Ratio) of whether a connection is established between two actors can be expressed as a form similar to the logit model.
\begin{equation}
\label{eq}
\log it(Y_{ij}=1)=\ln\left(\frac{p_{ij}}{1-p_{ij}}\right)=\theta ^T \delta [g(y,\mathbf{X})]_{ij}
\end{equation}
The regression coefficient of the model is similar to that of the logit model. The importance of the statistic $[g(y,\mathbf{X})]_{ij}$ to the connection probability is measured. The positive value represents an increasing effect of the connection formation, and the negative value indicates the less probability of the formation of connection. This paper estimates the model using statnet, ergm, network, sna and other packages in R software.
\\
\\

\section*{ Results}

\subsection*{ Index of Cross-regional Stars Cooperation}

\begin{figure}[!ht]
\centering
\includegraphics[width=.4\linewidth]{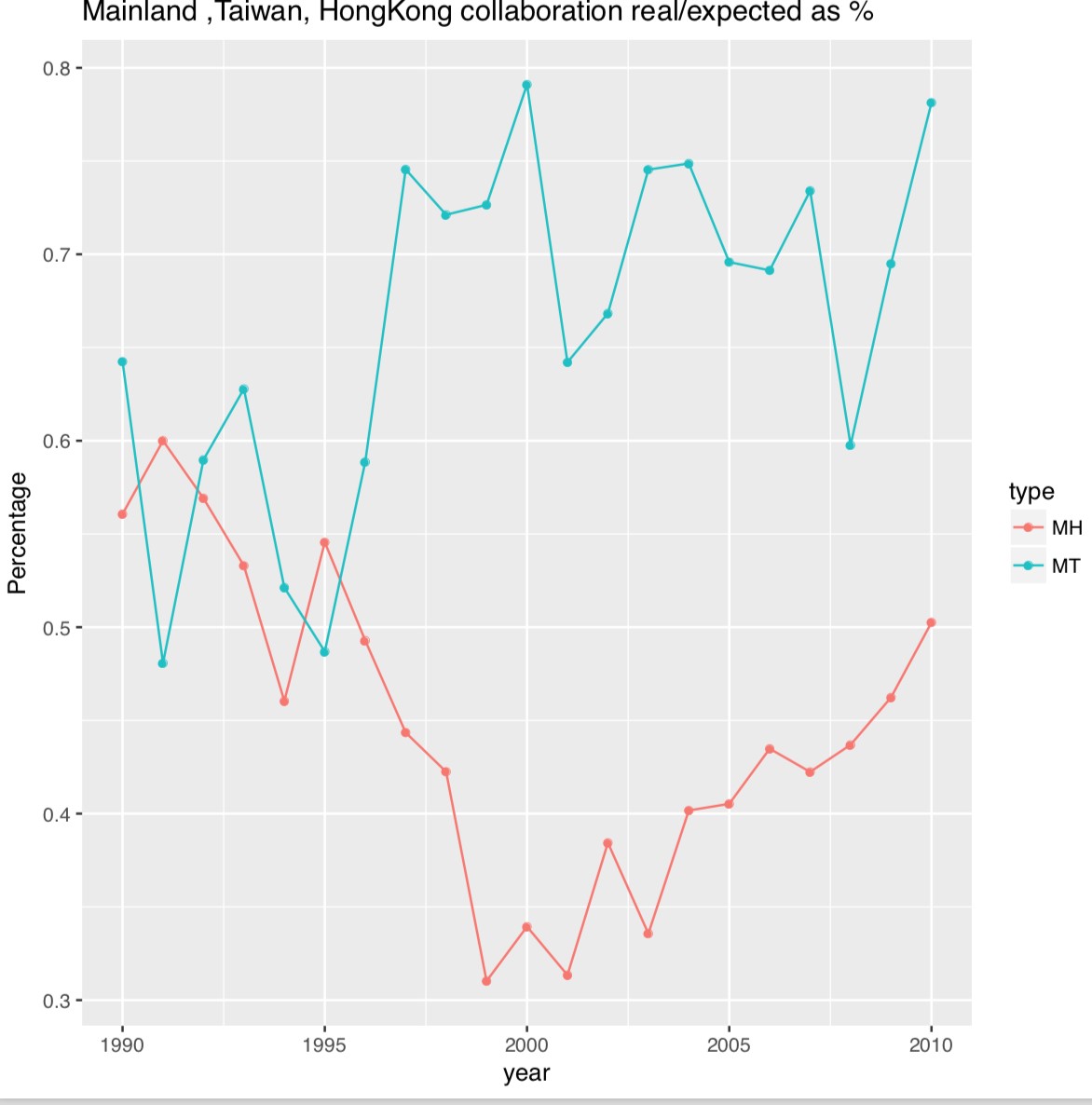}
\caption{The trends of the cross-regional cooperation indexes among Mainland-Hongkong and Mainland-Taiwan stars}
\label{fig3}
\end{figure}

\noindent Fig.~\ref{fig3} shows the changes of the index of cross-regional cooperation. The red line is the index of Mainland-Hongkong stars from 1990 to 2014, and the green one is the index of Mainland-Taiwan stars from 1990 to 2014. The cooperation index of Mainland-Hongkong stars had been declining since 1990, and hit the bottom in 1990 before it rose back and peaked in 2010.
\\
\\
\noindent The cooperation between Mainland and Taiwan stars shows an interesting fact. From 1990 to 1996, the cooperation index fluctuated from 0.5 to 0.6, and it abruptly rose above 0.73 in 1997. The index of Mainland-Taiwan stars was at its peak during 1997 to 2000 - which has been above 0.65 in recent years - a period when the cooperation index of Mainland-Hongkong stars declined rapidly. However, the index witnessed a drastic decline after 2000. Afterwards, the cooperation index of stars of these two areas has been around 0.5.

\noindent The changes in the cross-regional cooperation indexes among the stars between Mainland and Hongkong, Mainland and Taiwan show that the return of Hongkong had significant influence on the cross-regional cooperation. Before Hongkong’s Return, due to the ambiguous policies, the cooperation between stars in the two places was cautious, and cross-regional cooperation was significantly reduced. This effect reached its lowest point in Year 1999. After the return, the political environment becomes gradually clear, the cooperation index of stars between Mainland and Hongkong has increased year by year with the introduction of corresponding incentives and cooperative scheme in these two places. The cross-regional cooperation of Mainland-Taiwan stars has also been influenced by the return of Hongkong. The number of stars involved in Mainland-Taiwan cooperation showed significant growth during the time when the cross-regional cooperation of Mainland-Hongkong stars was declining before the return of Hongkong. This paper suggests that there is a " substitution effect ", that is, in the case of stable demand in the movie entertainment market, the cross-regional cooperation between Mainland and Hongkong stars has reduced due to the temporary uncertainty of the institutional environment, and correspondingly, the cross-regional cooperation between Mainland and Taiwan stars has enhanced.

\subsection*{ The Result of ERGMs}

\noindent The regional homogeneity in ERGMs is the interaction item of two nodes, and if the regional attributes of Mainland, Hongkong and Taiwan are taken into consideration, it will be very difficult to analyze the difference in the influence of regional homogeneity between Mainland-Hongkong and Mainland-Taiwan. In order to facilitate analysis, the Mainland-Hongkong stars and Mainland-Taiwan stars are separated to construct the model. In the previous model, Taiwan stars are excluded, and only the influences of the homogeneity of the regional attribute of Mainland and Hongkong stars are compared. For comparing the changes in the influence of homogeneity, this paper separates the five time periods to construct the model and compare the changes of coefficients. The analysis results of these models are shown in  Table~\ref{tab2} and \ref{tab3}.
\\
\\
To demonstrate the model results clearly, the coefficients of the homogeneity of regional attribute in Table~\ref{tab2} and \ref{tab3} are summarized and displayed in Fig.~\ref{fig4} below. The blue broken line in the figure shows that the cooperation among stars from Mainland and Hongkong experienced a decline before a rise. The influence of the homogeneity of the origins of stars on conjunctive probability has increased from 1.1 during 1990-1993 to 1.81 during 1998-2001, and it then kept dropping, and hit 1.59 during 2006-2009.Increased influence of homogeneity means that the probability of intra-regional cooperation is higher than that of inter-regional cooperation and accordingly, cross-regional cooperation is more difficult, and vice versa. Therefore, the cooperation probability among stars from Mainland and Hongkong experienced decline first before rise, which is consistent with the analysis results of cross-regional cooperation index.

\begin{figure}[!ht]
\centering
\includegraphics[width=.4\linewidth]{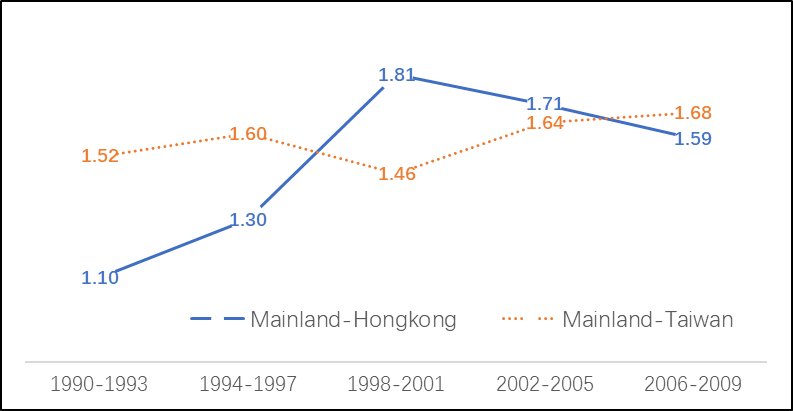}
\caption{Summary of ERGMs analysis results}
\label{fig4}
\end{figure}

\noindent The cooperation between Mainland and Taiwan, however, didn’t experience the same change. When other factors are controlled, the coefficient of the influence of regional attribute homogeneity between Mainland and Taiwan stars has been fluctuating between 1.5 and 1.7, shows a trend of decrease first and then increase, and reaches the lowest point between 1998 and 2001, that is, the inter-regional cooperation between Mainland-Taiwan stars is relatively easier and the effect of inter-regional cooperation between Mainland-Taiwan stars still exists in this period. The coefficient of regional homogeneity increased from 1.46 to 1.68 since 1998, indicating that inter-regional cooperation is becoming increasingly challenging, which is contrary to the situations between Mainland and Hongkong.
\begin{table}[!ht]
\centering
\begin{tabu}to \textwidth{X[-1lp]X[-1lp]X[-1lp]X[-1lp]X[-1lp]X[-1lp]}
\toprule
~& 
1990--1993& 
1994--1997& 
1998--2001& 
2002--2005& 
2006--2009 \\
\midrule
Edges Age Groups (Age group under 20 as reference)& 
$-$10.94 ***& 
$-$9.72 ***& 
$-$10.12 ***& 
$-$11.63 ***& 
$-$11.35 *** \\
20--39& 
0.36 ***& 
0.05 .& 
$-$0.1 ***& 
0.03 & 
0.01  \\
40--59& 
0.14 ***& 
$-$0.05 & 
$-$0.31 ***& 
0.03 & 
$-$0.02  \\
Over 60& 
0.11 & 
$-$0.15 *& 
$-$0.52 ***& 
$-$0.11 & 
$-$0.34 *** \\
Career Entry Time (Group before1980 as reference)& 
& 
& 
& 
& 
 \\
1980--1984& 
0.07 **& 
0.09 **& 
0.02 & 
0.1 ***& 
$-$0.1 *** \\
1985--1989& 
0.21 ***& 
0.03 & 
0.04 & 
0.00& 
$-$0.12 *** \\
1990--1994& 
0.66 ***& 
0.21 ***& 
0.07 *& 
0.08 & 
$-$0.15 *** \\
1995--1999& 
& 
$-$0.05 & 
0.28 ***& 
0.01 & 
$-$0.13 *** \\
2000--2004& 
& 
& 
$-$0.35 ***& 
0.39 ***& 
$-$0.03  \\
2005--2009& 
& 
& 
& 
$-$0.49 ***& 
0.47 *** \\
Stars Regional Attribute (Mainland as reference& 
0.34 ***& 
0.52 ***& 
0.37 ***& 
0.37 ***& 
0.45 *** \\
Number of cooperation in the previous period& 
0.34 ***& 
0.29 ***& 
0.33 ***& 
0.4 ***& 
0.41 *** \\
Inter-regional Cooperation in previous period& 
0.14 ***& 
$-$0.12 ***& 
$-$0.19 ***& 
0.01 & 
$-$0.11 *** \\
Homogeneity of Stars Regional Attribute& 
1.1 ***& 
1.3 ***& 
1.81 ***& 
1.71 ***& 
1.59 *** \\
AIC& 
65041 & 
87516 & 
134036 & 
228009 & 
321665  \\
BIC& 
65180 & 
87660 & 
134211 & 
228192 & 
321853  \\
Null Deviance& 
1052929 & 
1651348 & 
2672313 & 
4832788 & 
7131137  \\
Residual Deviance& 
65017 & 
87492 & 
134008 & 
227981 & 
321637  \\
\bottomrule
\end{tabu}
\caption{The ERGMs of cooperation between Mainland-Hongkong (1990--2009)}
\label{tab2}
\end{table}

\begin{table}[!ht]
\centering
\begin{tabu}to \textwidth{X[-1lp]X[-1lp]X[-1lp]X[-1lp]X[-1lp]X[-1lp]}
\toprule
~& 
1990--1993& 
1994--1997& 
1998--2001& 
2002--2005& 
2006--2009 \\
\midrule
Edges& 
$-$10.18 ***& 
$-$9.92 ***& 
$-$10.49 ***& 
$-$11.79 ***& 
$-$11.35 *** \\
Age Groups (Age group under 20 as reference) & 
& 
& 
& 
& 
 \\
20--39& 
0.31 ***& 
0.05 & 
$-$0.08 **& 
0.00& 
$-$0.01  \\
40--59& 
0.02 & 
$-$0.03 & 
$-$0.21 ***& 
$-$0.06 *& 
$-$0.04  \\
Over 60& 
$-$0.02 & 
$-$0.38 ***& 
$-$0.64 ***& 
$-$0.22 ***& 
$-$0.37 *** \\
Career Entry Time (Group before1980 as reference)& 
& 
& 
& 
& 
 \\
1980--1984& 
0.11 *& 
0.2 ***& 
0.17 ***& 
0.17 ***& 
$-$0.03  \\
1985--1989& 
0.06 & 
0.11 **& 
0.06 .& 
0.05 .& 
$-$0.09 *** \\
1990--1994& 
0.23 ***& 
0.2 ***& 
0.21 ***& 
0.14 ***& 
$-$0.13 *** \\
1995--1999& 
& 
$-$0.03 & 
0.39 ***& 
0.04 & 
$-$0.08 ** \\
2000--2004& 
& 
& 
$-$0.23 ***& 
0.45 ***& 
$-$0.01  \\
2005--2009& 
& 
& 
& 
$-$0.32 ***& 
0.52 *** \\
Stars Regional Attribute (Mainland as reference& 
0.65 ***& 
0.74 ***& 
0.46 ***& 
0.62 ***& 
0.57 *** \\
Number of cooperation in the previous period& 
0.25 ***& 
0.26 ***& 
0.32 ***& 
0.4 ***& 
0.39 *** \\
Inter-regional Cooperation in previous period& 
$-$0.2 ***& 
$-$0.25 ***& 
$-$0.01 & 
$-$0.07 ***& 
$-$0.09 *** \\
Homogeneity of Stars Regional Attribute& 
1.52 ***& 
1.6 ***& 
1.46 ***& 
1.64 ***& 
1.68 *** \\
AIC& 
34179 & 
53697 & 
111705 & 
213733 & 
319254  \\
BIC& 
34313 & 
53837 & 
111865 & 
213916 & 
319443  \\
Null Deviance& 
720443 & 
1199519 & 
2302293 & 
4763489 & 
7359713  \\
Residual Deviance& 
34155 & 
53673 & 
111679 & 
213705 & 
319226  \\
\bottomrule
\end{tabu}
\caption{The ERGMs of cooperation between Mainland-Taiwan (1990--2009)}
\label{tab3}
\end{table}

\section*{ Discussion}

\noindent This paper has obtained consistent results of the influence of homogeneity of regional attribute analyzed by both the cross-regional star cooperation index and ERGMs. Hongkong’s return exerts a significant influence on the cross-regional cooperation between Mainland-Hongkong stars, and the specific influence is divided into two parts: 1) the cross-regional cooperation between stars in the two places was cautious before Hongkong’s Return due to institutional uncertainty and ambiguous policy, the cross-regional cooperation decreased significantly in those years, and such decreasing trend continued until 1999; 2) after Hongkong’s Return, with the gradually clear political environment and the introduction of corresponding incentives and cooperative scheme in the two places, the star cooperation indexes of Mainland-Hongkong has increased year by year. As a control group at the same period, there is no significant change in the cooperation between Mainland and Taiwan stars before and after Hongkong’s Return. It is also found through analysis that the fluctuation in Mainland-Hongkong cross-regional cooperation caused by Hongkong’s Return has brought substitution effect to Mainland-Taiwan cross-regional cooperation, that is, the cross-regional cooperation between Mainland-Taiwan stars reaches the peak at the lowest point of cross-regional cooperation between Mainland-Hongkong stars.
\\
\\
\noindent As to the specific changes in the cross-regional cooperation between Hongkong and Mainland stars, since it is impossible to model the cross-regional association and the interpretation of the interaction between variables is complex, the above research cannot provide the answer. This paper jumps out of the perspective of network analysis and carries out analysis using the number of times of cross-regional cooperation between the nodes, as shown in Table~\ref{tab4} below.
\begin{table}[!ht]
\centering
\begin{tabu}to \textwidth{X[2cm]X[1cm]X[1cm]X[1cm]X[1cm]X[1cm]X[1cm]X[1cm]X[1cm]X[1cm]X[1cm]}
\toprule
& 
\multicolumn{2}{c}{All Stars} & 
\multicolumn{2}{c}{Famous Stars} & 
\multicolumn{2}{c}{Less Famous Stars} & 
\multicolumn{2}{p{2cm}}{Stars of older generation} & 
\multicolumn{2}{p{2cm}}{Stars of new generation}  \\
 & 
H& 
M& 
H& 
M& 
H& 
M& 
H& 
M& 
H& 
M \\
\midrule
Before 1990& 
4.79& 
2.39& 
& 
& 
& 
& 
6.45& 
3.84& 
& 
 \\
1990$-$1993& 
4.43& 
1.96& 
4.79& 
1.73& 
3.08& 
1.16& 
5.59& 
3.63& 
& 
 \\
1994$-$1997& 
4.19& 
1.62& 
4.85& 
1.75& 
2.82& 
1.10& 
5.05& 
3.39& 
1.65& 
0.38 \\
1998$-$2001& 
3.82& 
1.09& 
5.26& 
1.27& 
2.27& 
0.69& 
4.28& 
2.43& 
2.09& 
0.51 \\
2002$-$2005& 
6.12& 
1.23& 
9.23& 
1.14& 
3.38& 
1.00& 
7.47& 
2.49& 
4.12& 
0.85 \\
2006$-$2009& 
7.67& 
1.27& 
13.42& 
1.30& 
3.93& 
0.93& 
8.82& 
2.86& 
5.48& 
0.97 \\
Total& 
5.22& 
1.42& 
6.13& 
1.56& 
3.16& 
0.94& 
6.16& 
3.12& 
3.85& 
0.82 \\
\bottomrule
\end{tabu}
\caption{Number of Times of Cross-regional Cooperation between Hongkong and Mainland Stars over the Years}
\label{tab4}
\end{table}
\\
\\
\noindent By comparing the cross-regional cooperation of Hongkong stars in Table~\ref{tab4} with that of Taiwan stars in Table~\ref{tab5}, it is found that the number of times of cross-regional cooperation of Hongkong stars increased significantly after 1998.

\begin{table}[!ht]
\centering
\begin{tabu}to \textwidth{X[2cm]X[1cm]X[1cm]X[1cm]X[1cm]X[1cm]X[1cm]X[1cm]X[1cm]X[1cm]X[1cm]}
\toprule
& 
\multicolumn{2}{c}{All Stars} & 
\multicolumn{2}{c}{Famous Stars} & 
\multicolumn{2}{c}{Less Famous Stars} & 
\multicolumn{2}{p{2cm}}{Stars of older generation} & 
\multicolumn{2}{p{2cm}}{Stars of new generation}  \\
 & 
H& 
M& 
H& 
M& 
H& 
M& 
H& 
M& 
H& 
M \\
\midrule
Before 1990& 
4.00& 
1.60& 
~& 
~& 
~& 
~& 
6.30& 
2.59& 
~& 
~ \\
1990--1993& 
2.19& 
0.44& 
2.18& 
0.44& 
2.02& 
0.32& 
3.02& 
0.86& 
~& 
~ \\
1994--1997& 
2.72& 
0.52& 
2.04& 
0.58& 
2.86& 
0.40& 
2.53& 
1.02& 
0.89& 
0.19 \\
1998--2001& 
4.16& 
0.80& 
3.99& 
0.87& 
2.99& 
0.60& 
5.04& 
1.33& 
1.74& 
0.55 \\
2002--2005& 
5.00& 
0.98& 
5.57& 
0.92& 
2.71& 
0.86& 
7.48& 
1.29& 
2.70& 
0.86 \\
2006--2009& 
5.83& 
1.08& 
6.84& 
0.98& 
3.45& 
0.97& 
9.89& 
1.43& 
4.01& 
1.02 \\
Total& 
4.47& 
0.92& 
4.45& 
0.95& 
3.00& 
0.76& 
5.56& 
1.45& 
3.04& 
0.84 \\
\bottomrule
\end{tabu}
\caption{Number of Times of Cross-regional Cooperation between Taiwan and Mainland Stars over the Years}
\label{tab5}
\end{table}


\noindent \textbf{Research limitations:} In this paper, there is no distinction between movies and TV series; no distinction between different types of movies including mainland movies, Hong Kong movies or co-productions; the comparison of coefficient of ERGM model is problematic.


\begin{thebibliography}{31}

\bibitem{1}Newman David. 2006. "The lines that continue to separate us: borders in ourborderless' world"[J]. \emph{Progress in Human geography,}(2): 143-161.
\bibitem{2}Mccallum John. 1995. "National Borders Matter: Canada-U.S. Regional Trade Patterns"[J]. \emph{American Economic Review,}(3): 615-623.
\bibitem{3}Li Xun, Xu Xian xiang. 2006. "On the temporo-spatial variations of the border effects: approach and empirics"[J]. \emph{Geographical Research}(5): 792-802.
\bibitem{4}YANG Xiao-zhong and ZHANG Jie and YE Shu-juan. 2010. The Measure and Transformation of Border Effect of Cross-border Tourism Region Based on Social Network [J]. \emph{Scientia Geographica Sinica}(6): 826-832.
\bibitem{5}Helliwell, J. F. 1997. "National borders, trade and migration"[J]. \emph{Pacific Economic Review}, 2(3), 165-185.
\bibitem{6}Hazledine, T. 2009. "Border effects for domestic and international Canadian passenger air travel"[J]. \emph{Journal of Air Transport Management}, 15(1), 7-13.
\bibitem{7}YU Bin, LIU Ming-hua, ZHU Li-xia, et al. 2012. "The Border Effect and The Development of Border Region Between Conurbations"[J]. \emph{Scientia Geographica Sinica}, 32(6): 666-672.
\bibitem{8}Expert Paul, Evans Tim S., Blondel Vincent D., Lambiotte Renaud. 2011. "Uncovering space-independent communities in spatial networks"[J]. \emph {Proceedings of the National Academy of Sciences}(19): 7663-7668.
\bibitem{9}Grauwin Sebastian, Szell Michael, Sobolevsky Stanislav, Hövel Philipp, Simini Filippo, Vanhoof Maarten, Smoreda Zbigniew, Barabási Albert-László, Ratti Carlo. 2017. "Identifying and modeling the structural discontinuities of human interactions"[J]. \emph{Scientific reports}: 46677.
\bibitem{10}Hawelka Bartosz, Sitko Izabela, Beinat Euro, Sobolevsky Stanislav, Kazakopoulos Pavlos, Ratti Carlo. 2014. "Geo-located Twitter as proxy for global mobility patterns"[J]. \emph{Cartography and Geographic Information Science}(3): 260-271.
\bibitem{11}Ratti Carlo, Sobolevsky Stanislav, Calabrese Francesco, Andris Clio, Reades Jonathan, Martino Mauro, Claxton Rob, Strogatz Steven H. 2010. "Redrawing the map of Great Britain from a network of human interactions"[J]. \emph{PloS one}(12): e14248
\bibitem{12}Rinzivillo Salvatore, Mainardi Simone, Pezzoni Fabio, Coscia Michele, Pedreschi Dino, Giannotti Fosca. 2012. "Discovering the geographical borders of human mobility"[J]. \emph{KI-Künstliche Intelligenz}(3): 253-260.
\bibitem{13}Szell Michael, Sinatra Roberta, Petri Giovanni, Thurner Stefan, Latora Vito. 2012. "Understanding mobility in a social petri dish"[J].  \emph{Scientific reports}: 457.
\bibitem{14}Thiemann Christian, Theis Fabian, Grady Daniel, Brune Rafael, Brockmann Dirk. 2010. "The structure of borders in a small world"[J]. \emph{PloS one}(11): e15422.
\bibitem{15}Kang Chaogui, Sobolevsky Stanislav, Liu Yu, Ratti Carlo. 2013. "Exploring human movements in Singapore: a comparative analysis based on mobile phone and taxicab usages"[Z]. \emph{Proceedings of the 2nd ACM SIGKDD international workshop on urban computing}. ACM.
\bibitem{16}Liu Y, Sui Z, Kang C, et al. 2014. "Uncovering patterns of inter-urban trip and spatial interaction from social media check-in data"[J].  \emph{PloS one}, 9(1): e86026.
\bibitem{17}Shi L, Wu L, Chi G, et al. 2016. "Geographical impacts on social networks from perspectives of space and place: an empirical study using mobile phone data"[J]. \emph{Journal of Geographical Systems}, 18(4): 359-376.
\bibitem{18}Wang Z, Ye X, Lee J, et al. 2018. "A spatial econometric modeling of online social interactions using microblogs"[J]. \emph{Computers, Environment and Urban Systems}, 70: 53-58.
\bibitem{19}Goldenberg, J., \& Levy, M. 2009. "Distance is not dead: Social interaction and geographical distance in the internet era"[J]. \emph{arXiv preprint arXiv}:0906.3202.
\bibitem{20}Gao S, Liu Y, Wang Y, et al. 2013. "Discovering Spatial Interaction Communities from Mobile Phone Data"[J]. \emph{Transactions in GIS}, 17(3): 463-481
\bibitem{21}Cairncross Frances. 1997. "\emph{The death of distance}"[M]. Harvard Business School Press.
\bibitem{22}Ohmae Kenichi. 2015. "\emph{The borderless world}"[M]. China Citic Press.
\bibitem{23}Sparke, M., Sidaway, J. D., Bunnell, T., \& Grundy‐Warr, C. 2004. "Triangulating the borderless world: geographies of power in the Indonesia–Malaysia–Singapore growth triangle"[J]. \emph{Transactions of the Institute of British Geographers}, 29(4), 485-498.
\bibitem{24}Walther, O., \& Reitel, B. 2013. "ross-border policy networks in the Basel region: The effect of national borders and brokerage roles"[J]. \emph{Space and Polity}, 17(2), 217-236.
\bibitem{25}Newman M E J, Girvan M. 2004. "Finding and evaluating community structure in networks"[J]. \emph{Physical review E}, 69(2): 026113..
\bibitem{26}Frank O, Strauss D. 1986. "Markov graphs"[J]. \emph{Journal of the american Statistical association}, 81(395): 832-842.
\bibitem{27}Goodreau S M, Kitts J A, Morris M. 2009. "Birds of a feather, or friend of a friend? Using exponential random graph models to investigate adolescent social networks[J]". \emph{Demography}, 46(1): 103-125.
\bibitem{28}Wasserman S, Pattison P. 1996. "Logit models and logistic regressions for social networks: I. An introduction to Markov graphs and p"[J]. \emph{Psychometrika}, 61(3): 401-425.
\bibitem{29}Handcock, M. S., Robins, G., Snijders, T., Moody, J., \& Besag, J. 2003. "\emph{Assessing degeneracy in statistical models of social networks}"(Vol. 39). Working paper.
\bibitem{30}Snijders T A B. 2002. “Markov chain Monte Carlo estimation of exponential random graph models”[J]. \emph{Journal of Social Structure}, 3(2): 1-40.

\bibitem{31}Liu, D. and Albergante, L., 2017. "Balance of thrones: a network study on'Game of Thrones'." \emph{ preprint arXiv}:1707.05213.

\bibitem{32}Liu, D., Albergante, L. and Newman, T.J.2017 "Universal attenuators and their interactions with feedback loops in gene regulatory networks." "\emph{Nucleic acids research} 45.12 (2017): 7078-7093.

\bibitem{33}Liu, D., Davila-Velderrain, J., Zhang, Z. and Kellis, M.,2017 "Integrative construction of regulatory region networks in 127 human reference epigenomes by matrix factorization." \emph{bioRxiv}: 217588.




\end{thebibliography}
\end{document}